
\magnification 1200
\hsize 16. true cm
\baselineskip = 6truemm
%
\def\Z#1{\zeta(#1)}
\def\parn{\par\noindent}
\def\app{{\left({\alpha\over \pi}\right)}}
\def\PR{{\it Phys.Rev. }}
\def\PRL{{\it Phys.Rev.Lett. }}
\def\ref#1{[#1]}

\def\BPKud{{b+(k_1-k_2)^2}}

\def\Li{{\rm Li}}

\def\d${$ \displaystyle }

\def\DPKu{((p-k_1)^2+1)}
\def\DPKd{(p-k_2)^2  }

\def\DKue{(k_1^2  -i\epsilon)}
\def\DKde{(k_2^2+1-i\epsilon)}
\def\DKude{((k_1-k_2)^2+1-i\epsilon)}
\def\DPKue{((p-k_1)^2+1  -i\epsilon)}
\def\DPKde{((p-k_2)^2    -i\epsilon)}
\def\DQe{(q^2+1          -i\epsilon)}
\def\DQue{((q-k_1)^2+1   -i\epsilon)}
\def\DQde{((q-k_2)^2     -i\epsilon)}

\def\oo{\infty}

\def\LS{l}
\def\MS{m}
\def\DLS{d\LS}
\def\DMS{d\MS}

\def\RBLM{\;R(b,-\LS,-\MS)\;}
\def\ROOT{\sqrt{\LS(\LS+4)}}

\def\DLS{d\LS}
\def\DMS{d\MS}

\def\A#1{a_#1}
\def\LGd{\ln 2}
\def\LGdd{\ln^2 2}

\def\LGdq{\ln^4 2}

\def\Graph{{\rm graph \;}}
\def\ll{\ln{\lambda}}
\def\lld{\ln^2{\lambda}}
\def\d${$ \displaystyle }

\rightline{\bf DFUB 94-18}
\rightline{21 September 1994}
\vskip 20truemm
\centerline{\bf The analytical value of the corner-ladder graphs}
\par
\centerline{\bf contribution to the electron ($g$-$2$) in QED. }
\par
\vskip 20truemm
\centerline{S.Laporta
\footnote {$^{\S}$}{ {E-mail: {\tt laporta@bo.infn.it} }}  }
\vskip 20truemm
\centerline{\it Dipartimento di Fisica, Universit\`a di Bologna,}
\centerline{\it and INFN, Sezione di Bologna,}
\centerline{\it Via Irnerio 46, I-40126 Bologna, Italy}
\vskip 20truemm

\centerline{\bf Abstract }  \par
The contributions to the ($g$-$2$) of the electron from the
corner-ladder graphs in sixth-order (three-loop) QED perturbation theory
are evaluated in closed analytical form.
The results obtained are in excellent agreement with the most precise
numerical evaluations already existing in the literature.
Our results allows one to reduce the numerical uncertainty
of the theoretical determination of the $g$-$2$ of the electron.
\vskip 2truemm
\vfill\eject

We have calculated in closed analytical form the contribution
to the anomalous magnetic moment of the electron
at sixth-order (three-loop) in QED perturbation theory
from the so-called corner-ladder graphs shown in Fig.1.

The result are, accounting for the mirror graphs
(the infrared terms proportional to $\ll$ and $\lld$ are omitted,
they are listed in ref.[1]):

$$ \eqalign{
     a(\Graph  7)=
       -& {5\over 2}\Z5 + {5\over 9}\pi^2\Z3
        + {47\over 720}\pi^4 + 2\pi^2\LGdd - 28\A4  - {7\over 6}\LGdq
        - {80\over 9}\Z3
        \  \cr &
        + {55\over 18}\pi^2\LGd
        - {2743\over 1296}\pi^2     + {2521\over 864}
                                                     \ ,\cr }  \eqno(1) $$
$$ \eqalign{
     a(\Graph  9)=
       &  {95\over 24}\Z5 - {43\over 72}\pi^2\Z3
        + {199\over 2160}\pi^4  + {1\over 9}\pi^2\LGdd  - {44\over 3}\A4
                                                   - {11\over 18}\LGdq
        \qquad \quad \cr &
        - {107\over 12}\Z3  + {25\over 18}\pi^2\LGd
        + {3\over 16}\pi^2  + {43\over 36}
                                                 \ ,\cr}  \eqno(2) $$
$$ \eqalign{
     a(\Graph 17)=
       &  {25\over 6}\Z5 - {2\over 3}\pi^2\Z3
        - {31\over 540}\pi^4 - {113\over 54}\pi^2\LGdd + {200\over 9}\A4
                              + {25\over 27}\LGdq
        \qquad \cr &
       + {199\over 24}\Z3 - {155\over 18}\pi^2\LGd
         + {3809\over 648}\pi^2 - {29\over 27}
                                                  \ ,\cr} \eqno(3) $$
$$ \eqalign{
     a(\Graph 19)=
       &  {95\over 24}\Z5 - {3\over 8}\pi^2\Z3
        - {43\over 432}\pi^4  + {37\over 18}\pi^2\LGdd - {28\over 3}\A4
                               - {7\over 18}\LGdq
        \qquad \quad \cr &
        - {635\over 72}\Z3 + {83\over 18}\pi^2\LGd \
        - {4777\over 2592}\pi^2 + {1835\over 864}
                                                  \ , \cr}  \eqno(4) $$
$$ \eqalign{
     a(\Graph 27)=
       -& {215\over 24}\Z5  + {95\over 72}\pi^2\Z3
        + {41\over 180}\pi^4 - {137\over 27}\pi^2\LGdd + {160\over 9}\A4
                               + {20\over 27}\LGdq
        \cr &
        + {69\over 4}\Z3  - {101\over 18}\pi^2\LGd
         + {2401\over 2592}\pi^2 - {3017\over 864}
                                                 \ ,\cr  } \eqno(5) $$
$$ \eqalign{
     a({\rm total})=
       &  {5\over 8}\Z5  + {17\over 72}\pi^2\Z3
        + {493\over 2160}\pi^4 - 3\pi^2\LGdd - 12\A4  - {1\over 2}\LGdq
        - {13\over 12}\Z3
        \quad \cr &
        - {31\over 6}\pi^2\LGd
        + {655\over 216}\pi^2 + {481\over 288}
                                                  \ . \cr} \eqno(6) $$
Here and in the following $\zeta(p)$ is the Riemann
$\zeta$-function of argument $p$,
\d$ \zeta(p) \equiv \sum_{n=1}^{\infty} {1\over n^p} \ ,$
(we recall that
$\zeta(2)=\pi^2/6$,
$\zeta(3)=1.202\;056\;903...$,
$\zeta(4)=\pi^4/90$,
$\zeta(5)= 1.036\;927\;755...$), and
\d$ a_4 \equiv \sum_{n=1}^{\infty} {1\over {2^n n^4}}
   = 0.517\;479\;061...\ .$

The contributions of these graphs had been previously calculated
only by (approximate) numerical methods. The numerical values of
eqs.(1)-(6) turn out to be in excellent agreement with
the values shown in ref.[1]  and [2] (see Table I).

If $c_3$ is the coefficient of the \d$\app^3$ term in the perturbative
 expansion of the electron anomaly in QED
$$ a_e({\rm QED})= {1\over 2} \app
+ c_2 \app^2 +c_3\app^3 +c_4\app^4 + ...\ , \eqno(7) $$
by using eq.(6) we have
$$ c_3 = 1.176\;19(21) \ , \eqno(8)$$
to be compared with the value [3]
$$ c_3 = 1.176\;13(42) \ ; \eqno(9)$$
the error of eq.(8) is due to numerical uncertainty of the value
of the contribution of the remaining group of five graphs still not
known in analytical form [4].
\parn
Using the experimental determination of $\alpha$ [5],
$$ \alpha^{-1}({\rm exp})=137.035\;997\;9(32) \ ,  \eqno(10) $$
the values of $c_2$ and $c_4$ [6]
$$c_2 = {197 \over 144} +{1\over 2 }\Z2 +{3 \over 4} \Z3
          -3 \Z2 \LGd =-0.328\;478\;965 ...$$
$$ c_4= -1.434(138) \eqno(11) $$
and accounting for the small vacuum polarization contributions
of muon, tau and hadron loops as well as the electroweak contribution
 [6]
one finds
$$ a_e({\rm theory})=  1\;159\;652\;141.4(27.1)(2.6)(4.1)\times 10^{-12}
\eqno(12)$$
where the first error comes from the error of the determination of
$\alpha$ (10), the second from the value of $c_3$ and the third
from the error of $c_4$.
The error due to the three-loop coefficient $c_3$ is now less
than the error due to $c_4$.

If one assumes the validity of QED,
using the experimental value of the electron anomaly [7]
$$a_e({\rm exp})= 1\;159\;652\;188.4 (4.3)\times 10^{-12} \ , \eqno(13)$$
one can estimate $\alpha$ from eq.(7);
one finds
$$ \alpha^{-1}({a_e})=137.035\;992\;34(51)(31)(48)  \ ,\eqno(14) $$
where the errors come respectively, from eq.(13), from $c_3$ and from
$c_4$.
Summing up the errors of eq.(14) in quadrature one finds
$$ \alpha^{-1}({a_e})=137.035\;992\;34(77)  \ \eqno(15) $$
which is the most accurate determination of $\alpha$ available
at present.

We describe now briefly the techniques used for deriving eqs.(1)-(6).
All the vertex graphs of Fig.1 can be obtained from the self-mass--like graph
of Fig.2 by inserting the external photon line in all possible ways along the
electron line.
Once that the limit $\Delta \to 0 $ is taken, the contribution
from each of graphs of Fig.1 is expressed as a sum of some hundreds
of terms
all of the form \d$N/D$, where $N$ is a scalar polynomial
in $p$ and the internal integration momenta,
and
\d$ D = D_1^{n_1}\ D_2^{n_2}\ ...\ D_8^{n_8}\ (n_i= 0,1,2),\ $
the $D_i$ being the denominators of the various propagators of the graph
of Fig.2.
The simplest term, with $N=1$ and all $n_i=1$, was calculated in
analytical form in a precedent paper [8].

The contributions of graphs 19 and 27 turn out to be finite;
on the contrary, graphs 7, 9 and 17 need renormalization,
and as consequence their contributions become infrared divergent;
in this case
appropriate ultraviolet counterterms are subtracted, as well as
infrared ones which render the entire contributions finite.

Let us consider for instance the term defined as
$$ I={1\over \pi^6} \int (-i)^3 d^4q \; d^4k_1 \; d^4k_2 \; {N\over D} \ ,
\eqno(16) $$
with $N=(k_1-k_2)^2$ and
\parn
\vskip 2 truemm
\parn
\d$ D= \DKue\DKde\DKude\DPKue\DPKde\times $
$$ \times \DQe\DQue\DQde $$
(the electron mass is set equal to 1).
In order to evaluate eq.(16) we apply the techniques
used for calculating the analytical contribution to the $g$-$2$ of the
muon of the ``light-light'' vertex graphs [9].
The self-mass--like graph of Fig.2 is
depicted so as  to show
the topological similarity with the correspondent self-mass--like graph
obtained from the ``light-light'' graphs.
Following [9], we write a dispersion relation in $(k_1-k_2)^2$ at constant
 $k_1^2=\LS$,
$k_2^2=\MS,$ for the inserted  $q$-loop vertex part
\parn
\vskip 2truemm
\d$ \int { (-i){d^4q}\over {\DQe\DQue\DQde} } = $
$$ =\pi^2
  \int^\oo_{1} { db \over \BPKud} {1 \over \RBLM } \ln{W} \ , $$
where
$$ \ln {W} \equiv \ln {
     {b(b+\LS+\MS)+(b-\LS+\MS)+(b-1)\;\RBLM} \over
     {b(b+\LS+\MS)+(b-\LS+\MS)-(b-1)\;\RBLM} } \ , \eqno(18)$$
and $R(x,y,z)$ is the usual two-body phase space square root
$$R(x,y,z)\equiv\sqrt{(x-y-z)^2-4yz}=\sqrt{x^2+y^2+z^2-2xy-2yz-2xz}
                                                       \ . \eqno(19)  $$
Now we introduce hyperspherical variables for both the
$k_1$ and the $k_2$ loops
$$ \eqalign{
& d^4k_1= {1\over 2} \LS \ \DLS \ d\Omega_4(\hat k_1) \ , \cr
& d^4k_2= {1\over 2} \MS \ \DMS \ d\Omega_4(\hat k_2) \ , \cr
    }  \eqno(20)$$
and perform the hyperspherical angular integrals
by expanding the denominators in Gegenbauer polynomials [10].
In the mass-shell limit $p^2=-1$ we obtain
$$ \int{ {d\Omega_4(\hat k_1)} {d\Omega_4(\hat k_2)}
          \over {\DPKu\DPKd(\BPKud)} } =
   { \left(2 \pi^2\right)^2 \over {\LS\MS}  }\ln Y_\pm \ ,  \eqno (21)$$
where
$$ Y_\pm \equiv     1+ {1\over 8\MS\LS }
                     \left[ \ROOT  -\LS \right]
                     \left[ (\MS+1) \pm(\MS-1) \right]
                     \left[b +\LS+\MS-\RBLM \right]  \  .$$
The analytical continuation for timelike values of $p^2$
of the radial integral over $\MS$
requires also
a deformation of the contour [10] of the integration
in order to avoid the singularity at $\MS=-1$.
The contour is split in two parts,
one where $\MS\ge -1$ and another where $-1\le \MS \le 0$.
$Y_{-}$ must be used when integrating over $\MS$ in the first zone and
and $Y_{+}$ in the second one.
At this point the integral (16) reads
$$ I =      \hskip -3pt
            \int^\oo_0        {d\LS\over \LS}
            \left(
            \int^\oo_{-1}     {d\MS\over {\MS+1}}
            \int^\oo_{1}      { db \over \RBLM } \ln{W} \ln {Y_{-}}
            \qquad\qquad\qquad\quad
            \right. $$
$$          \left.
	    -
            \int^0_{-1}      {d\MS\over {\MS+1}}
            \int^\oo_{1}      { db \over \RBLM } \ln{W} \ln {Y_{+}}
	    \right)                                    	    \  .
                                                         \eqno (22) $$
Eq.(22) contains two square roots: $\RBLM$ and $\ROOT$.
We choose to integrate eq.(22) in $b$ first, then in $\MS$
and at last in $\LS$; we used this order of integration
for all the other integrals over $b$, $\MS$ and $\LS$ needed in the
calculation of eqs.(1)-(6)
\footnote {$^{1}$}{
Note the simplification with respect to the analogous ``light-light''
graphs calculation;
in eq.(23) of ref.[9]
four square roots appear, and as consequence
the decomposition of the hyperspherical logarithm is needed
and different orders of integration must be chosen for the different
pieces.
}.
If we write
$$ I=\int^\oo_0 {d\LS\over\LS} f(l)\ , $$
the $b$ and $\MS$ integrals are performed by differentiating
(repeatedly, when needed) with respect to $\LS$
(see ref.[8]),
then integrating the resulting (simpler) expression, and recovering
$f(l)$ by a quadrature.
Note that the derivatives with respect to $\LS,\MS$ and $b$
of the functions $\ln W$ and $\ln Y_{\pm}$
contain the same polynomial in the denominator,
$(b+\MS)^2+\LS(b-1)(\MS+1)$,
even if the two functions have different origin;
this fact simplifies considerably the calculations.
At last, $f(l)$ is expressed as a sum of known tetralogarithms
of the variable
\d$ x=(\LS+2-\ROOT)/2 $ :
$$\eqalign{
  f(\LS)=&-9 \Li_4(x) -15 \Li_4(-x) +{1\over 2} S_{2,2}(x^2)
          -2 S_{2,2}(-x) +2 T_{2,2}(x)
          -\Li_2^2(x)                    \cr
        & +\ln x \left( 5\Li_3(x) +8 \Li_3(-x) -2 S_{1,2}(x)
          -2 \Li_2(x)\ln(1-x) \right)       \cr
        & -\ln^2 x \left( \Li_2(x) +{3\over 2} \Li_2(-x)
          +{1\over 2}\ln^2(1-x) \right)
         +2\Z2\Li_2(-x) \ ,} \eqno(23)    $$
where
$$ S_{n,p}(x)\equiv {(-1)^{n+p-1}\over {(n-1)! \; p!}} \int^1_0{
  {dt\over  t} \ln ^{n-1}(t)  \ln ^{p}(1-xt)       } \ , $$
$$ \Li_{n+1} (x)\equiv S_{n,1}(x) \ ,$$
and
\footnote {$^{2}$}{
The expression of the function $T_{2,2}(x)$
in terms of polylogarithmic functions $S_{n,p}(x)$ is not known;
nevertheless, the properties of $T_{2,2}(x)$ are similar to those
of $S_{n,p}(x)$, so that we consider this function as a ``standard''
polylogarithm.
}
$$ T_{2,2}(x)\equiv \int^x_0{
  {dt\over  t} \ln (t+1) \Li_2(t)} \ . $$
The last definite integral over $\LS$ is performed using known
tables of integrals containing polylogarithms;
one finds
$$ I= {21\over 2}\Z2\Z3 -{45\over 4}\Z5 \ .
 \eqno(24) $$

The processing of the other terms with different $N$ and $D$
is similar to that described above.
When the numerator $N$ contains powers of $q$ we use the decomposition
described in ref.[9]; as this fact causes a severe blowing-up of the
size of the calculations, some algebraic manipulations are used in order
to reduce the maximum power to $q^3$.
In the course of the calculations,
some non-standard functions of polylogarithmic degree 4
appear, as an example
$$ g_1(x)= \int^x_0 dt \ln(1-t+t^2) {\Li_2(t)\over t} \ ;$$
the expressions of the functions $g_i(x)$
by means of polylogarithmic functions $S_{n,p}(x)$ are not known.
The last integration over $\LS$ of the functions $g_i(x)$,
after some integration by parts,
give a group of definite integrals containing $\ln(1-t+t^2)$
which we found to be the same ones appeared in the calculation of the
contribution to the $g$-$2$ of the electron
of another set of vertex graphs [11],
and which were calculated in analytical form in ref.[12].

The final infrared-cutoff dependent results (1)-(6) are obtained
using the analytical expressions of the
infrared counterterms shown in ref.[13].

As a check, for each graph of Fig.1
we have modified the internal routing of the momentum
$\Delta$ and we have extracted the contribution to the ($g$-$2$),
obtaining new expressions, whose
subsequent analytical evaluations
reproduce correctly eqs.(1)-(6).

The algebra of the whole calculation was processed relying on the algebraic
manipulation program ASHMEDAI [14].
All calculations were done on a number of VAXstation 4000/60 and 4000/90.
The processing of the whole contribution of a graph required
about 40h of continued CPU time
and 150 Mbytes of disk space
for storing of intermediate expressions on a VAXstation 4000/90,
the peak length of the algebraic expressions being about half million
of terms.
Due to the (relative) high speed of machines used,
we devoted our efforts to design and carry out the maximum quantity
of consistency and cross checks, rather than to speed up the
execution time or to minimize the disk space requirements.

Work at the contribution to the electron anomaly of the remaining
graphs still not known in analytical form is now in progress.

\vskip 6 truemm
\parn
{\bf Acknowledgement}
\par
The author wants to thank E. Remiddi for continuous discussions and
encouragement and M. J. Levine for kindly providing
his program for algebraic manipulation ASHMEDAI.

\vfill\eject
\parn
{\bf References}
\vskip 10 truemm
\parn
\item{[1]} M. J. Levine, J.Wright, \PR D{\bf 8} (1973) 3171.
\parn
\item{[2]} M. J. Levine, H. Y. Park. and R. Z. Roskies,
                                         \PR D {\bf 25} (1982) 2205.
\parn
\item{[3]} S. Laporta and E. Remiddi, {\it Phys.Lett.} {\bf B} 265 (1991) 182.
\item{[4]} T. Kinoshita
            in {\it Quantum Electrodynamics},
            edited by T.Kinoshita,
            Advanced series on Directions in High Energy Physics, Vol. 7,
            (World Scientific, Singapore, 1990), p.218.
\parn
\item{[5]} M. E. Cage {\it et al.}, {\it IEEE Trans. Instrum. Meas.},
            {\bf 38} (1989) 233.
\item{[6]} T. Kinoshita and W. B. Lindquist, \PR D {\bf 42} (1990) 636.
\item{[7]} R. S. Van Dick, Jr., P. B. Schwinberg and H. G. Dehmelt,
                    \PRL  {\bf 59} (1987) 26.
\item{[8]} M. Caffo, E. Remiddi and S. Turrini,
              {\it Nuovo Cimento A}, {\bf 79} (1984) 220.
\parn
\item{[9]} S. Laporta and E. Remiddi, {\it Phys.Lett.} {\bf B} 301 (1993) 440.
\parn
\item{[10]} M. J. Levine and R. Roskies,
                                     \PR D{\bf 9}, (1974) 421.
\parn
\item{[11]} M. J. Levine, E. Remiddi and R. Roskies, \PR D {\bf 20} (1979)
2068.
\parn
\item{[12]} S. Laporta, \PR D {\bf 47} (1993) 4793.
\parn
\item{[13]} M. J. Levine, E. Remiddi and R. Roskies,
            in {\it Quantum Electrodynamics},
            edited by T.Kinoshita,
            Advanced series on Directions in High Energy Physics, Vol. 7,
            (World Scientific, Singapore, 1990), p.176.
\parn
\item{[14]} M. J. Levine, U.S. AEC Report No. CAR-882-25 (1971),
                                                     unpublished.
\parn
\vfill\eject
\parn
\centerline {\bf Figure captions}
\parn
\bigskip
\item{Fig.1:} The corner-ladder graphs (mirror graphs are omitted).
              The numbering follows ref.[1].
\parn
\item{Fig.2:} The self-mass--like scalar graph with the cut of the
dispersion relation.
\parn
\vfill\eject
\parn
--------------------------------------------------------
--------------------------------------------------------
\parn
\centerline {Table I}
--------------------------------------------------------
--------------------------------------------------------
\parn
\hskip 10 truemm       Graph
\hskip 17 truemm       ref.[1]
\hskip 20 truemm       ref.[2]
\hskip 28 truemm       Our
$$
\eqalign{
& 7    \qquad   \qquad   \cr
& 9       \cr
& 17      \cr
& 19      \cr
& 27      \cr
& {\rm Total}   \cr}
\eqalign{
-&2.6707(19) \qquad    \cr
&0.6189(64)   \cr
&0.6097(34)   \cr
-&0.3182(72)  \cr
&1.8572(86)   \cr
&0.0893(60)   \cr }
\eqalign{
-&2.670\;546(30) \qquad    \cr
&0.617\;727(121) \cr
&0.607\;660(240) \cr
-&0.334\;698(11) \cr
&1.861\;992(240) \cr
&0.082\;065(362) \cr }
\eqalign{
-&2.670\;554\;745\;651... \qquad    \cr
&0.617\;711\;782\;178...  \cr
&0.607\;752\;806\;792...  \cr
-&0.334\;695\;103\;723... \cr
&1.861\;907\;872\;591...  \cr
&0.082\;122\;612\;187...  \cr }
$$
\parn
--------------------------------------------------------
--------------------------------------------------------
\parn
Table I: Comparison of our values of the contributions
of the graphs shown in Fig.1 with the values of ref.[1] and ref.[2].
\parn
\bye